\documentclass[twocolumn,showpacs,prx,superscriptaddress,floatfix]{revtex4}

\usepackage{color}
\usepackage{tabularx}
\usepackage{epsfig}
\usepackage{amsmath}
\usepackage{amssymb}
\usepackage{graphicx}
\usepackage{wasysym}
\usepackage{dcolumn}
\usepackage{bm}
\usepackage{subfigure}
\usepackage{psfrag}
\usepackage{subfigure}
\usepackage{times}

\begin{document}

\title{Erratum: Glassy Chimeras Could Be Blind to Quantum Speedup\ldots
[Phys.~Rev.~X {\bf 4}, 021008 (2014)]}

\author{Martin Weigel}
\affiliation{Applied Mathematics Research Centre, Coventry University,
Coventry, CV1 5FB, United Kingdom}

\author{Helmut G.~Katzgraber}
\affiliation{Department of Physics and Astronomy, Texas A\&M University,
College Station, Texas 77843-4242, USA}
\affiliation{Materials Science and Engineering Program, Texas A\&M
University, College Station, Texas 77843, USA}
\affiliation{Santa Fe Institute, 1399 Hyde Park Road, Santa Fe, New Mexico 87501, USA}

\author{Jonathan Machta}
\affiliation{Physics Department, University of Massachusetts, Amherst,
Massachusetts 01003 USA}
\affiliation{Santa Fe Institute, 1399 Hyde Park Road, Santa Fe, New Mexico 87501, USA}

\author{Firas Hamze}
\affiliation {D-Wave Systems, Inc., 3033 Beta Avenue, Burnaby, British
Columbia, V5G 4M9, Canada}

\author{Ruben S.~Andrist}
\affiliation{Santa Fe Institute, 1399 Hyde Park Road, Santa Fe, NM 87501, USA}

\collaboration{The Octomore Collaboration}

\date{\today}


\maketitle

\begin{figure}[h]
\includegraphics[width=0.90\columnwidth]{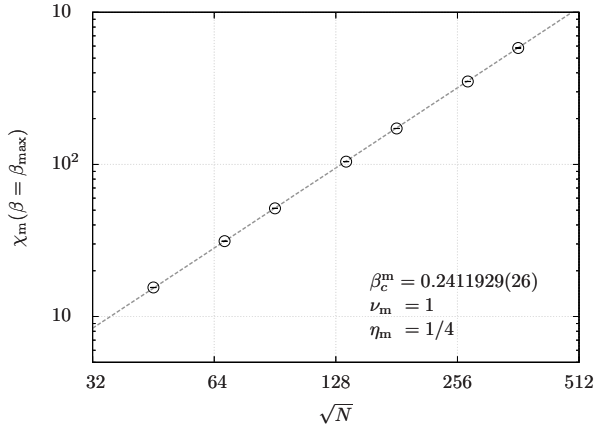}
\vspace*{-1.0em}
\caption{ 
Log-log plot of the FM susceptibility at its peaks $\chi_{\rm m}(\beta
= \beta_{\mathrm{max}})$ as a function of the effective linear system
size $\sqrt{N}$. The data fall onto a straight line, showing that
corrections to scaling are very small. The dashed line has slope $2 -
\eta = 7/4$.}
\label{fig:chi}
\end{figure}

In Ref.~\onlinecite{katzgraber:14} (2014), Katzgraber {\em et
al.}~studied Ising spins on the Chimera topology \cite{bunyk:14} of the
D-Wave Two quantum annealing machine both in the spin-glass (SG) as well
as the ferromagnetic (FM) sector.  For the simulations in the FM sector,
the following critical parameters based on simulations of systems of up
to $N = 3200$ spins were computed: $\beta_c^{\rm m} = 0.2402(3)$,
$\nu_{\rm m} \approx 1$, and $\eta_{\rm m} \approx 2/5$.  Here,
$\beta_c^{\rm m} = 1/T_c^{\rm m}$ is the inverse critical temperature,
$\nu^{\rm m}$ the critical exponent of the FM correlation length, and
$\eta^{\rm m}$ the critical exponent associated with the FM
susceptibility.  Using single-cluster updates \cite{wolff:89}, Monte
Carlo simulations of up to $N= 131\,072$ spins in the FM sector have
been performed. A detailed finite-size scaling analysis of the specific
heat, the magnetization, the susceptibility, and the Binder parameter
and related cumulants, as well as three logarithmic derivatives of the
magnetization together with an analysis of cross-correlations
\cite{weigel:09,weigel:10} leads to the following estimates of the
critical parameters: $\beta_c^{\rm m} = 0.241\,1929(26)$, $\nu_{\rm m} =
0.9980(15)$, and $\eta_{\rm m} = 0.2513(14)$.  Hence, the critical
exponents of a FM Ising model on the Chimera lattice are in agreement
with the exact values for the two-dimensional (2D) Ising model, namely
$\nu_{\rm 2D} = 1$ and $\eta_{\rm 2D} = 1/4$ \cite{yeomans:92}.
Therefore, both models share the same universality class
\cite{comment:peter}.  Figure \ref{fig:chi} illustrates the perfect
agreement between our new simulations and the exact values. Plotted are
the FM susceptibility at its maximum, i.e., for $\beta =
\beta_\mathrm{max}(N)$, vs the number of spins, where $\chi_{\rm
m}(\beta = \beta_\mathrm{max}) \sim (\sqrt{N})^{2-\eta_{\rm m}}$.
Indeed, the dashed line has slope $2 - \eta_{\rm 2D} = 7/4$ and is a
guide to the eye.  We would like to conclude by emphasizing the
following points.

\vspace*{-0.5em}

\begin{enumerate} \itemsep0pt 

\item{For the FM sector, only the critical exponent $\eta_{\rm m}$ is
incorrect in Ref.~\cite{katzgraber:14}, as well as the claim that the
universality class of FM Ising models in 2D and on the Chimera topology
might be different.}

\item{For the SG sector $\eta_{\rm q} = 0$ or close to zero is
plausible.  However, simulations to verify this would be difficult.}

\item{The main conclusions of Ref.~\cite{katzgraber:14} remain
unchanged.}

\item{We would like to caution researchers seeking quantum speedup based
on systems with $512$ qubits or less, because, while corrections to
scaling are weak, they seem to persist up to large system sizes.}

\end{enumerate}

\vspace*{-0.5em}

\noindent {\em Acknowledgments} --- M.~W.~and H.~G.~K.~acknowledge support
from the European Commission through the IRSES network DIONICOS under
Contract No.~PIRSES-GA-2013-612707.  H.G.K.~acknowledges support from
the NSF (Grant No.~DMR-1151387) and would like to thank Bruichladdich
Dist.~for providing continued inspiration.  J.~M.~acknowledges support
from the NSF (Grant No.~DMR-1208046).

\bibliography{refs,comments}

\begin{thebibliography}{7}
\bibitem[{com()}]{comment:peter}
\bibinfo{note}{Yes, Peter, you were right.}
\end{thebibliography}

\end{document}